\documentclass{article}
\usepackage{amssymb,amsmath,amsthm}
\newcommand{\coloneq}{\mathrel{\texttt{:\!-}}}
\newcommand{\dcont}{\subseteq^{\Delta}}

\newtheorem{proposition}{Proposition}

\theoremstyle{definition}
\newtheorem{definition}{Definition}
\newtheorem{example}{Example}

\title{Relational Association Rules: \\ getting {\sc Warm{\rm e}r}}
\author{Bart Goethals and Jan Van den Bussche \\ University of Limburg, Belgium}
\date{}

\begin{document}
\maketitle

\begin{abstract}
In recent years, the problem of association rule mining in
transactional data has been well studied.
We propose to extend the discovery of classical
association rules to the discovery of association rules
of conjunctive queries in arbitrary relational data,
inspired by the {\sc Warmr} algorithm, developed by Dehaspe and Toivonen,
that discovers association rules over a limited set of conjunctive queries.
Conjunctive query evaluation in relational databases is well
understood, but still poses some great challenges when approached
from a discovery viewpoint in which patterns are generated
and evaluated with respect to some well defined search space and pruning operators.
\end{abstract}

\section{Introduction}

In recent years, the problem of mining association rules over
frequent itemsets in transactional data~\cite{dmprinc} has been
well studied and resulted in several algorithms that can find
association rules within a limited amount of time. Also more
complex patterns have been considered such as
trees~\cite{zakitrees}, graphs~\cite{subgraphicdm,subgraphpkdd},
or arbitrary relational structures~\cite{warmr,warmr2}. However,
the presented algorithms only work on databases consisting of a
set of transactions.  For example, in the tree
case~\cite{zakitrees}, every transaction in the database is a
separate tree, and the presented algorithm tries to find all
frequent subtrees occurring within all such transactions.
Nevertheless, many relational databases are not suited to be
converted into a transactional format and even if this were
possible, a lot of information implicitly encoded in the
relational model would be lost after conversion. Towards the
discovery of association rules in arbitrary relational databases,
Deshaspe and Toivonen developed an inductive logic programming
algorithm, {\sc Warmr}~\cite{warmr,warmr2}, that discovers
association rules over a limited set of conjunctive queries on
transactional relational databases in which every transaction
consists of a small relational database itself. In this paper, we
propose to extend their framework to a broader range of
conjunctive queries on arbitrary relational databases.

Conjunctive query evaluation in relational databases is well
understood, but still poses some great challenges when approached
from a discovery viewpoint in which patterns are generated and
evaluated with respect to some well defined search space and
pruning operators. We describe the problems occurring in this
mining problem and present an algorithm that uses a similar
two-phase architecture as the standard association rule mining
algorithm over frequent itemsets (Apriori)~\cite{kddboek_chap12},
which is also used in the {\sc Warmr} algorithm. In the first
phase, all frequent patterns are generated, but now, a pattern is
a conjunctive query and its support equals the number of distinct
tuples in the answer of the query. The second phase generates all
association rules over these patterns. Both phases are based on
the general levelwise pattern mining algorithm as described by
Mannila and Toivonen~\cite{border}.

In Section~\ref{problem}, we formally state the problem we try to
solve. In Section~\ref{general}, we describe the general approach
that is used for a large family of data mining problems. In
Section~\ref{warmr}, we describe the {\sc Warmr} algorithm which
is also based on this general approach. In Section~\ref{warmer},
we describe our approach as an generalization of the {\sc Warmr}
algorithm and identify the algorithmic challenges that need to be
conquered. In Section~\ref{sample}, we show a sample run of the
presented approach. We conclude the paper in
Section~\ref{conclusion} with a brief discussion and future work.

\section{Problem statement} \label{problem}

The relational data model is based on the idea of representing
data in tabular form.  The \emph{schema} of a relational database
describes the names of the tables and their respective sets of
column names, also called attributes.  The actual content of a
database, is called an \emph{instance} for that schema. In order
to retrieve data from the database, several query languages have
been developed, of which SQL is the standard adopted by most
database management system vendors.  Nevertheless, an important
and well-studied subset of SQL, is the family of conjunctive
queries.

As already mentioned in the Introduction, current algorithms for
the discovery of patterns and rules mainly focused on
transactional databases.  In practice, these algorithms use
several specialized data structures and indexing schemes to
efficiently find their specific type of patterns, i.e., itemsets,
trees, graphs, and many others.  As an appropriate generalization
of these kinds of patterns, we propose a framework for arbitrary
relational databases in which \emph{a pattern is a conjunctive
query}.

Assume we are given a relational database consisting of a schema
${\bf R}$ and an instance ${\bf I}$ of ${\bf R}$. An \emph{atomic
formula} over ${\bf R}$ is an expression of the form $R(\bar x)$,
where $R$ is a relation name in ${\bf R}$ and $\bar x$ is a
$k$-tuple of variables and constants, with $k$ the arity of $R$.

\begin{definition}
A \emph{conjunctive query} $Q$ over ${\bf R}$ consists of a \emph{head}
and a \emph{body}. The body is a finite set of atomic formulas over ${\bf R}$.
The head is a tuple of variables occurring in the body.
\end{definition}

A \emph{valuation} on $Q$ is a function $f$ that assigns a constant
to every variable in the query.
A valuation is a \emph{matching} of $Q$ in ${\bf I}$, if for every $R(\bar x)$
in the body of $Q$, the tuple $f(\bar x)$ is in ${\bf I}(R)$.
The \emph{answer} of $Q$ on ${\bf I}$ is the set
$$Q({\bf I}) := \{f(\bar y)\mid \bar y \mbox{ is the head of } Q \mbox{ and } f \mbox{ is a matching of } Q \mbox{ on } {\bf I} \}.$$

We will write conjunctive queries using the commonly used Prolog notation.
For example, consider the following query on a beer drinkers database:
$$ Q(x) \coloneq {\it likes}(x,\text{`Duvel'}), {\it likes}(x,\text{`Trappist'}). $$
The answer of this query consists of all drinkers that like Duvel and also like Trappist.

For two conjunctive queries $Q_1$ and $Q_2$ over ${\bf R}$,
we write $Q_1 \subseteq Q_2$ if for every possible instance ${\bf I}$ of ${\bf R}$,
$Q_1({\bf I}) \subseteq Q_2({\bf I})$ and say that $Q_1$ is \emph{contained} in $Q_2$.
$Q_1$ and $Q_2$ are called \emph{equivalent}
if and only if $Q_1 \subseteq Q_2$ and $Q_2 \subseteq Q_1$.
Note that the question whether a conjunctive query is contained in another conjunctive query
is decidable~\cite{ullmanvol2}.

\begin{definition}
The \emph{support} of a conjunctive query $Q$ in an
instance ${\bf I}$ is the number of distinct
tuples in the answer of $Q$ on ${\bf I}$. A query is called
\emph{frequent} in ${\bf I}$ if its support exceeds a given
\emph{minimal support threshold}.
\end{definition}

\begin{definition}
An \emph{association rule} is of the form $Q_1 \Rightarrow Q_2$, such
that $Q_1$ and $Q_2$ are both conjunctive queries and $Q_2 \subseteq Q_1$.
An association rule is called \emph{frequent} in ${\bf I}$ if $Q_2$ is
frequent in ${\bf I}$ and it is called \emph{confident} if the support of $Q_2$
divided by the support of $Q_1$ exceeds a given \emph{minimal confidence} threshold.
\end{definition}

\begin{example}
Consider the following two queries:
\begin{align*}
& Q_1(x,y) \coloneq {\it likes}(x,\text{`Duvel'}), {\it visits}(x,y). \\
& Q_2(x,y) \coloneq {\it likes}(x,\text{`Duvel'}), {\it visits}(x,y), {\it serves}(y,\text{`Duvel'}).
\end{align*}
The rule $Q_1 \Rightarrow Q_2$ should then be read as follows:
if a person $x$ that likes Duvel visits bar $y$, then bar $y$ serves Duvel.
\end{example}

A natural question to ask is why we should only consider rules over queries
that are contained for any possible instance.
For example, assume we have the following two queries:
\begin{align*}
& Q_1(y) \coloneq {\it likes}(x,\text{`Duvel'}), {\it visits}(x,y). \\
& Q_2(y) \coloneq {\it serves}(y,\text{`Duvel'}).
\end{align*}
Obviously, $Q_2$ is not contained in $Q_1$ and vice versa.
Nevertheless, it is still possible that for a given instance ${\bf I}$,
we have $Q_2({\bf I}) \subseteq Q_1({\bf I})$, and hence this could make
an interesting association rule $Q_1 \Rightarrow Q_2$,
which should be read as follows:
if bar $y$ has a visitor that likes Duvel, then bar $y$ also serves Duvel.
\begin{proposition} \label{whycontain}
Every association rule $Q_1 \Rightarrow Q_2$, such that $Q_2({\bf I}) \subseteq Q_1({\bf I})$,
can be expressed by an association rule $Q_1 \Rightarrow Q_2'$,
with $Q_2' = Q_2 \cap Q_1$, and essentially has the same meaning.
\end{proposition}
In this case the correct rule would be $Q_1 \Rightarrow Q_2$, with
\begin{align*}
& Q_1(y) \coloneq {\it likes}(x,\text{`Duvel'}), {\it visits}(x,y). \\
& Q_2(y) \coloneq {\it likes}(x,\text{`Duvel'}), {\it visits}(x,y), {\it serves}(y,\text{`Duvel'}).
\end{align*}
Note the resemblance with the queries used in Example~1.
The bodies of the queries are the same, but now we have another head.
Evidently, different heads result in a different meaning of the corresponding
association rule which can still be interesting.
As another example, note the difference with the following two queries:
\begin{align*}
& Q_1(x) \coloneq {\it likes}(x,\text{`Duvel'}), {\it visits}(x,y). \\
& Q_2(x) \coloneq {\it likes}(x,\text{`Duvel'}), {\it visits}(x,y), {\it serves}(y,\text{`Duvel'}).
\end{align*}
The rule $Q_1 \Rightarrow Q_2$ should then be read as follows:
if a person $x$ that likes Duvel visits a bar, then $x$ also visits a bar that serves Duvel.

The goal is now to find all frequent and confident association
rules in the given database.

\section{General approach} \label{general}

As already mentioned in the introduction,
most association rule mining algorithms use the common two-phase architecture.
Phase 1 generates all frequent patterns,
and phase 2 generates all frequent and confident association rules.

The algorithms used in both phases are based on the general
levelwise pattern mining algorithm as described by Mannila and
Toivonen~\cite{border}. Given a database ${\cal D}$, a class of
patterns ${\cal L}$, and a selection predicate $q$, the algorithm
finds the ``theory'' of ${\cal D}$ with respect to ${\cal L}$ and
$q$, i.e.,  the set ${\cal T}\!h({\cal L}, {\cal D}, q) := \{\phi
\in {\cal L} \mid q({\cal D},\phi) \mbox{ is true}\}$. The
selection predicate $q$ is used for evaluating whether a pattern
$Q \in {\cal L}$ defines a (potentially) interesting pattern in
${\cal D}$. The main problem this algorithm tries to tackle is to
minimize the number of patterns that need to be evaluated by $q$,
since it is assumed this evaluation is the most costly operation
of such mining algorithms. The algorithm is based on a
breadth-first search in the search space spanned by a
specialization relation which is a partial order $\preceq$ on the
patterns in ${\cal L}$. We say that $\phi$ is \emph{more specific}
than $\psi$, or $\psi$ is \emph{more general} than $\phi$, if
$\phi \preceq \psi$. The relation $\preceq$ is a \emph{monotone
specialization relation} with respect to $q$, if the selection
predicate $q$ is monotone with respect to $\preceq$, i.e.,  for
all ${\cal D}$ and $\phi$, we have the following: if $q({\cal
D},\phi)$ and $\phi \preceq \gamma$, then $q({\cal D},\gamma)$. In
what follows, we assume that $\preceq$ is a monotone
specialization relation. We write $\phi \prec \psi$ if $\phi
\preceq \psi$ and not $\psi \preceq \phi$. The algorithm works
iteratively, alternating between \emph{candidate generation} and
\emph{candidate evaluation}, as follows.
\begin{tabbing}
$C_1:=\{ \phi \in {\cal L} \mid \mbox{there is no } \gamma \mbox{ in } {\cal L}
\mbox{ such that } \phi \prec \gamma \};$ \\
$i := 1;$ \\
\textbf{while} $C_i \neq \emptyset$ \textbf{do} \\
\quad \= // Candidate evaluation \\
\> ${\cal F}_i := \{\phi \in C_i \mid q({\cal D},\phi)\};$ \\
\> // Candidate generation \\
\> $C_{i+1} := \{\phi \in {\cal L} \mid \mbox{for all } \gamma, \mbox{ such that }\phi \prec \gamma,
 \mbox{ we have } \gamma \in \bigcup_{j \leq i} {\cal F}_j\} \backslash \bigcup_{j \leq i} C_j ;$\\
\> $i:=i+1$ \\
\textbf{end while} \\
\textbf{return} $\bigcup_{j < i} {\cal F}_j$;
\end{tabbing}
In the generation step of iteration $i$, a collection $C_{i+1}$ of new \emph{candidate patterns}
is generated, using the information available from the more general patterns in $\bigcup_{j \leq i} {\cal F}_j$,
which have already been evaluated.
Then, the selection predicate is evaluated on these candidate patterns.
The collection ${\cal F}_{i+1}$ will consist of those patterns in $C_{i+1}$ that satisfy the selection predicate $q$.
The algorithm starts by constructing $C_1$ to contain all most general patterns.
The iteration stops when no more potentially interesting patterns can be found
with respect to the selection predicate.

In general, given a language ${\cal L}$ from which patterns are
chosen, a selection predicate $q$ and a monotone specialization
relation $\preceq$ with respect to $q$, this algorithm poses
several challenges.
\begin{enumerate}

\item An initial set $C_1$ of most general candidate patterns needs to be identified,
which is not always possible for infinite languages, and hence other, maybe less
optimal solutions could be required.

\item Given all patterns $\bigcup_{j \leq i} {\cal F}_j$ that satisfy the selection predicate
up to a certain level $i$, the set $C_{i+1}$ of all candidate patterns must be generated efficiently.
It might be impossible to generate all but only those elements in $C_{i+1}$,
but instead, it might be necessary to generate a superset of $C_{i+1}$ after which
the non candidate patterns must be identified and removed.
Even if this identification is efficient, naively generating all possible patterns could
still become infeasible if this number of patterns becomes too large.
Hence, this poses two additional challenges:
\begin{enumerate}
\item efficiently generate the smallest possible superset of $C_{i+1}$, and
\item identify and remove each generated pattern that is no candidate pattern
by efficiently checking whether all of its generalizations are in $\bigcup_{j \leq i} {\cal F}_j$.
\end{enumerate}

\item Extract all patterns from $C_{i+1}$ that satisfy the selection predicate $q$,
by efficiently evaluating $q$ on all elements in $C_{i+1}$.

\end{enumerate}

In the next section, we identify these challenges for both phases of
the association rule mining problem within the framework
proposed by Dehaspe and Toivonen, and describe their solutions
as implemented within the {\sc Warmr} algorithm.

\section{The {\sc Warmr} algorithm} \label{warmr}

As already mentioned in the introduction, a first approach
towards the goal of discovering all frequent and confident association
rules in arbitrary relational databases, has been presented by Dehaspe and Toivonen,
in the form of an inductive logic programming algorithm,
{\sc Warmr}~\cite{warmr,warmr2}, that discovers association rules
over a limited set of conjunctive queries.

\subsection{Phase 1}

The procedure to generate all frequent conjunctive queries is primarily based on
a \emph{declarative language bias} to constrain the
search space to a subset of all conjunctive queries,
which is an extensively studied subfield in ILP\@~\cite{ilp}.

The declarative language bias used in {\sc Warmr} drastically simplifies
the search space of all queries by using the {\sc Warmode} formalism.
This formalism requires two major constraints.
The most important constraint is the \emph{key constraint}.
This constraint requires the specification of a single \emph{key} atomic formula
which is obligatory in all queries.
This key atomic formula also determines \emph{what} is counted, i.e.,  it
determines the head of the query, that is, all variables occuring in the key atom.
Second, it requires a list $\textit{Atoms}$ of all atomic formulas that are allowed in
the queries that will be generated.  In the most general case, this list consists
of the relation names in the database schema ${\bf R}$.
If one also wants to allow certain constants within the
atomic formulas, then these atomic formulas must be specified for every such constant.
In the most general case, the complete database instance must also be added to the $\textit{Atoms}$ list.
The {\sc Warmode} formalism also allows other constraints,
but since these are not obligatory, we will not discuss them any further.
\begin{example}
Consider
\begin{align*}
\textit{Atoms} := \{ & \textit{likes}(\_,\text{`Duvel'}), \\
& \textit{likes}(\_,\text{`Trappist'}), \\
& \textit{serves}(\_,\text{`Duvel'}), \\
& \textit{serves}(\_,\text{`Trappist'})\},
\end{align*}
where $\_$ stands for an
arbitrary variable, and $$\textit{key} := \textit{visits}(\_,\_).$$
Then,
\begin{align*}
{\cal L} = \{ & Q(x_1,x_2) \coloneq \textit{visits}(x_1,x_2), \textit{likes}(x_3,\text{`Duvel'}). \\
& Q(x_1,x_2) \coloneq \textit{visits}(x_1,x_2), \textit{likes}(x_1,\text{`Duvel'}). \\
& \ldots \\
& Q(x_1,x_2) \coloneq \textit{visits}(x_1,x_2), \textit{serves}(x_3,\text{`Duvel'}). \\
& Q(x_1,x_2) \coloneq \textit{visits}(x_1,x_2), \textit{serves}(x_2,\text{`Duvel'}). \\
& \ldots \\
& Q(x_1,x_2) \coloneq \textit{visits}(x_1,x_2), \textit{likes}(x_1,\text{`Duvel'}), \textit{serves}(x_2,\text{`Duvel'}). \\
& Q(x_1,x_2) \coloneq \textit{visits}(x_1,x_2), \textit{likes}(x_1,\text{`Duvel'}), \textit{serves}(x_2,\text{`Trappist'}). \\
& \ldots \}.
\end{align*}
\end{example}
As can be seen, these constraints already dismiss a lot
of interesting patterns.
However, it is still possible to discover all frequent conjunctive queries,
but then, we need to run the algorithm for every possible key atomic formula
with the least restrictive declarative language bias.
Of course, using this strategy, a lot of possible optimizations are left
out, as will be shown in the next section.

The specialization relation used in {\sc Warmr} is defined $Q_1 \preceq Q_2$ if $Q_1 \subseteq Q_2$.
The selection predicate $q$ is the minimal support threshold, which is indeed monotone
with respect to $\preceq$,
i.e.,  for every instance ${\bf I}$ and conjunctive queries $Q_1$ and $Q_2$,
we have the following: if $Q_1$ is frequent and $Q_1 \subseteq Q_2$,
then $Q_2$ is frequent.

\paragraph{Candidate generation}
In essence, the {\sc Warmr} algorithm generates all
conjunctive queries contained in the query $Q({\bar x}) \coloneq R({\bar x})$,
where $R({\bar x})$ is the key atomic formula. Denote this query by the \emph{key conjunctive query}.
Hence, the key conjunctive query is the (single) most general pattern in $C_1$.
Assume we are given all frequent patterns up to a certain level $i$, $\bigcup_{j \leq i} {\cal F}_j$.
Then, {\sc Warmr} generates a superset of all candidate patterns,
by adding a single atomic formula, from $\textit{Atoms}$, to every query in ${\cal F}_i$,
as allowed by the {\sc Warmode} declarations.
From this set, every candidate pattern needs to be identified by checking whether
all of its generalizations are frequent.
However, this is no longer possible, since some of these generalizations might
not be in the language of admissible patterns.
Therefore, only those generalizations that satisfy the declarative language bias
need to be known frequent.  In order to do this, for each generated query $Q$,
{\sc Warmr} scans all infrequent conjunctive queries for one that is more general than $Q$.
However, this does not imply that all queries that are more general than $Q$ are known to be frequent!
Indeed, consider the following example which is based on the declarative language bias from the previous example.
\begin{example} \label{fout}
\begin{align*}
& Q_1(x_1,x_2) \coloneq {\it visits}(x_1,x_2), {\it likes}(x_1,\text{`Duvel'}). \\
& Q_2(x_1,x_2) \coloneq {\it visits}(x_1,x_2), {\it likes}(x_3,\text{`Duvel'}).
\end{align*}
Both queries are single extensions of the key conjunctive query, and hence, they are generated within
the same iteration.  Obviously, $Q_2$ is more general than $Q_1$, but still, both queries
remain in the set of candidate queries.  Moreover, it is necessary that both queries remain admissible,
in order to guarantee that all frequent conjunctive queries are generated.
\end{example}
This example shows that the candidate generation step of {\sc
Warmr} does not comply with the general levelwise framework given
in the previous section. Indeed, at a certain iteration, it
generates patterns of different levels in the search space spanned
by the containment relation.

The generation strategy also generates several queries that are equivalent with other candidate queries,
or with queries already generated in previous iterations,
which also need to be identified and removed from the set of candidate patterns.
Again, for each candidate query, all other candidate queries and all frequent queries are
scanned for an equivalent query.
Unfortunately, the question whether two conjunctive queries are equivalent
is an NP-complete problem.
Note that isomorphic queries are definitely equivalent (but not vice versa in general),
and also the problem of efficiently generating finite structures up to isomorphism,
or testing isomorphism of two given finite structures efficiently, is still an
open problem~\cite{graphisomorph}.

\paragraph{Candidate evaluation}
Since {\sc Warmr} is an inductive logic programming algorithm
written within a logic programming environment,
the evaluation of all candidate queries is performed inefficiently.
Still, {\sc Warmr} uses several optimizations to increase the
performance of this evaluation step, but these optimizations
can hardly be compared to the optimized
query processing capabilities of relational database systems.

\subsection{Phase 2}

The procedure to generate all association rules in {\sc Warmr},
simply consists of finding all couples $(Q_1,Q_2)$ in the list of frequent queries,
such that $Q_2$ is contained in $Q_1$.
We were unable to find how this procedure exactly works,
that is, how is each query $Q_2$ found, given query $Q_1$.
Anyhow, in general, this phase is less of an efficiency issue,
since the supports of all queries that need to be considered are already known.

\section{Getting {\sc Warm{\rm e}r}} \label{warmer}

Inspired by the framework of {\sc Warmr}, we present in this
section a more general framework and investigate the efficiency
challenges described in Section~\ref{general}. More specifically,
we want to discover association rules over all conjunctive queries
instead of only those queries contained in a given key conjunctive
query since it might not always be clear what exactly needs to be
counted. For example, in the beer drinkers database, the examples
given in section~\ref{problem} show that different heads could
lead to several interesting association rules about the drinkers,
the bars or the beers separately. We also want to exploit the
containment relationship of conjunctive queries as much as
possible, and avoid situations such as described in
example~\ref{fout}. Indeed, the {\sc Warmr} algorithm does not
fully exploit the different levels induced by the containment
relationship, since it generates several candidate patterns of
different levels within the same iteration.

\subsection{Phase 1}

The goal of this first phase is to find all frequent conjunctive queries.
Hence, ${\cal L}$ is the family of all conjunctive queries.

Since only the number of different tuples in the answer of a query
is important and not the content of the answer itself,
we will extend the notion of query containment,
such that it can be better exploited in the levelwise algorithm.

\begin{definition}
A conjunctive query $Q_1$ is \emph{diagonally contained} in $Q_2$ if
$Q_1$ is contained in a projection of $Q_2$.
We write $Q_1 \dcont Q_2$.
\end{definition}

\begin{example}
\begin{align*}
& Q_1(x) \coloneq {\it likes}(x,y), {\it visits}(x,z), {\it serves}(z,y) \\
& Q_2(x,z) \coloneq {\it likes}(x,y), {\it visits}(x,z), {\it serves}(z,y)
\end{align*}
The answer of $Q_1$ consists of all drinkers that visit at least one bar that serve
at least one beer they like.  The answer of $Q_2$ consists of all visits of a drinker
to a bar if that bar serves at least one beer the drinker likes.
Obviously, a drinker could visit multiple bars that serve a beer they like, and hence
all these bars will be in the answer of $Q_2$ together with that drinker, while
$Q_1$ only gives the name of that drinker, and hence, the number of tuples in the answer
of $Q_1$ will always be smaller or equal than the number of tuples in the answer of $Q_2$.
\end{example}

We now define $Q_1 \preceq Q_2$ if $Q_1 \dcont Q_2$. The selection
predicate $q$ is the minimal support threshold, which is indeed
monotone with respect to $\preceq$, i.e.,  for every instance
${\bf I}$ and conjunctive queries $Q_1$ and $Q_2$, we have the
following: if $Q_1$ is frequent and $Q_1 \dcont Q_2$, then $Q_2$
is frequent. Notice that the notion of diagonal containment now
allows the incorporation of conjunctive queries with different
heads within the search space spanned by this specialization
relation.

Two issues remain to be solved:
how are the candidate queries efficiently generated
without generating two equivalent queries? and how is the frequency of
each candidate query efficiently computed?

\paragraph{Candidate generation}
As a first optimization towards the generation of all
conjunctive queries, we will already prune several queries
in advance.
\begin{enumerate}
\item The head of a query must contain at least one variable,
since the support of a query with an empty head can be at most 1.
Hence, we already know its support after we evaluate a query with
the same body but a nonempty head.

\item We allow only a single permutation of the head,
since the supports of queries with an equal body but different
permutations of the head are equal.
\end{enumerate}

Generating candidate conjunctive queries using the levelwise algorithm
requires an initial set of all most general queries with respect to $\dcont$.
However, such queries do not exist. Indeed, for every conjunctive query $Q$,
we can construct another conjunctive query $Q'$, such that $Q \dcont Q'$ by
simply adding a new atomic formula with new variables into the body of $Q$,
and adding these variables to the head.
A rather drastic but still reasonable solution to this problem is to apriori
limit the search space to conjunctive queries with at most a fixed
number of atomic formulas in the body.
Then, within this space, we can look at the set of most general queries,
and this set now is well-defined.

At every iteration in the levelwise algorithm we need to generate all candidate conjunctive queries
up to equivalence, such that all of their generalizations are known to be frequent.
Since an algorithm to generate exactly this set is not known,
we will generate a small superset of all candidates and afterwards remove each
query of which a generalization is not known to be frequent (or known to be infrequent).

Nevertheless, any candidate conjunctive query is always more specific than
at least one query in ${\cal F}_i$.
Hence, we can generate a superset of all possible candidate queries
using the following four operations on each query in ${\cal F}_i$.
\begin{description}
\item[Extension:] We add a new atomic formula with new variables to the body.
\item[Join:] We replace all occurrences  of a variable with another variable already occurring in the query.
\item[Selection:] We replace all occurrences of a variable $x$ with some constant.
\item[Projection:] We remove a variable from the head if this does not result in an empty head.
\end{description}
\begin{example}
This example shows a single application of each operation on the query
$$ Q(x,y) \coloneq {\it likes}(x,y),{\it visits}(x,z), {\it serves}(z,u).$$
\begin{description}
\item[Extension:] $$ Q(x,y) \coloneq {\it likes}(x,y),{\it visits}(x,z),{\it serves}(z,u),{\it likes}(v,w).$$
\item[Join:] $$ Q(x,y) \coloneq {\it likes}(x,y), {\it visits}(x,z), {\it serves}(z,y).$$
\item[Selection:] $$ Q(x,y) \coloneq {\it likes}(x,y), {\it visits}(x,z), {\it serves}(z,\text{`Duvel'}).$$
\item[Projection:] $$ Q(x) \coloneq {\it likes}(x,y), {\it visits}(x,z), {\it serves}(z,u).$$
\end{description}
\end{example}
The following proposition implies that if we apply a sequence of these four operations on the current
set of frequent conjunctive queries, we indeed get at least all candidate queries.
\begin{proposition}
\emph{$Q_1 \dcont Q_2$ if and only if a query equivalent to $Q_1$ can be obtained from $Q_2$ by
applying some finite sequence of extension, join, selection and projection operations.}
\end{proposition}
Nevertheless, using these operations, several equivalent or redundant queries can be generated.
An efficient algorithm avoiding the generation of equivalent queries
is still unknown.  Hence, whenever we generate a candidate query,
we need to test whether it is equivalent with another query we already generated.
In order to keep the generated superset of all candidate conjunctive queries as small
as possible, we apply an operator once on each query.  If the query is redundant
or equivalent with a previously generated query, we repeatedly apply any of the operators
until a query is found that is not equivalent with a previously generated query.
As already mentioned in the previous section, testing equivalence cannot be done efficiently.

After generating this superset of all candidate conjunctive queries,
we need to check for each of them whether all more general conjunctive queries
are known to be frequent.
This can be done by performing the inverses of the four operations extension, join, selection
and projection, as described above.
Even if we now assume that in the set of all frequent conjunctive queries
there exist no two equivalent queries, we still need to find
the query equivalent to the one generated using the inverse operations.
Hence, the challenge of testing equivalence of two conjunctive queries reappears.

\paragraph{Candidate evaluation}
After generating all candidate conjunctive queries,
we need to test which of them are frequent.
This can be done by simply evaluating every candidate query on the database,
one at a time, by translating each query to SQL.
Although conjunctive query evaluation in relational databases is well
understood and several efficient algorithms have been developed
(i.e.,  join query optimisation and processing)~\cite{Ullmanimpl},
this remains a costly operation.
Within database research, a lot of research has been done on multi-query optimization~\cite{multiquery}.
Here, one tries to efficiently evaluate multiple queries at once.
Unfortunately, these techniques are not yet implemented in most common database systems.

As a first optimization towards query evaluation, we can already
derive the support of a significant part of all candidate conjunctive queries.
Therefore, we only consider those candidate queries that satisfy the following restrictions.
\begin{enumerate}
\item We only consider queries that have no constants in the head,
because the support of such queries is equal to the support
of those queries in which the constant is not in the head.

\item We only consider queries that contain no
duplicate variables in the head, since the support of such a query
is equal to the support of the query without duplicates in the head.

\end{enumerate}

As another optimization, given a query involving constants,
we will not treat every variation of that query that uses
different constants as a separate query,
but rather we can evaluate all those variations in a single global query.
For example, suppose the query $$ Q(x_1) \coloneq R(x_1,x_2) $$ is frequent.
From this query, a lot of candidate queries are generated using the
selection operation on $x_2$. Assume the active domain of $x_2$ is
$1,2,\ldots,n$, then the set of candidate queries contains at least
$$ \{Q(x_1) \coloneq R(x_1,1), Q(x_1) \coloneq R(x_1,2), \ldots,
Q(x_1) \coloneq R(x_1,n) \}, $$
resulting in a possibly huge amount of queries that need to be evaluated.
However, the support of all these queries can be computed
by evaluating only the single SQL query
\begin{tabbing}
\qquad \= {\bf select} $x_2$, {\bf count}$(*)$ \\
\>{\bf from} $R$ \\
\>{\bf group by} $x_2$ \\
\>{\bf having count}$(*) \geq \textit{minsup}$
\end{tabbing}
of which the answer consists of every possible constant $c$ for $x_2$ together
with the support of the corresponding query $Q(x_1) \coloneq R(x_1,c)$.
From now on, we will therefore use only a symbolic constant to denote
all possible selections of a given variable.
For example, $Q(x_1) \coloneq R(x_1,c_1)$ denotes the set of all
possible selections for $x_2$ in the previous example.
A query with such a symbolic constant is then considered frequent
if it is frequent for at least one constant.

As can be seen, several optimizations can be used to improve
the performance of the evaluation step in our algorithm.
Also, we might be able to use some of the techniques
that have been developed for frequent itemset mining,
such as closed frequent itemsets~\cite{closed}, free sets~\cite{dfree}
and non derivable itemsets~\cite{ndi}.
These techniques could then be used to minimize the number of candidate queries
that need to be executed on the database, but instead we might be able to compute
their supports based on the support of previously evaluated queries.
Another interesting optimization could be to avoid using SQL queries completely,
but instead use a more intelligent counting mechanism that needs to scan the database
or the materialized tables only once, and count the supports of all queries at the same time.

\subsection{Phase 2}

The goal of the second phase is to find for every frequent conjunctive
query $Q$, all confident association rules $Q' \Rightarrow Q$.
Hence, we need to run the general levelwise algorithm separately
for every frequent query.
That is, for any given $Q$, ${\cal L}$ consists of all conjunctive queries $Q'$,
such that $Q \subseteq Q'$.
Assume we are given two association rules $\textit{AR}_1: Q_1 \Rightarrow Q_2$
and $\textit{AR}_2: Q_3 \Rightarrow Q_4$, we define
$\textit{AR}_1 \preceq \textit{AR}_2$ if $Q_3 \subseteq Q_1$ and $Q_2 \subseteq Q_4$.

The selection predicate $q$ is the minimal confidence threshold which is
again monotone with respect to $\preceq$,
i.e., for every instance ${\bf I}$ and association rules $\textit{AR}_1: Q_1 \Rightarrow Q_2$
and $\textit{AR}_2: Q_3 \Rightarrow Q_4$, we have the following: if $\textit{AR}_1$ is frequent and confident
and $\textit{AR}_1 \preceq \textit{AR}_2$, then $\textit{AR}_2$ is frequent and confident.

Here, only a single issue remains to be solved:
how are the candidate queries efficiently generated
without generating two equivalent queries?

We have to generate, for every frequent conjunctive query $Q$,
all conjunctive queries $Q'$, such that $Q \subseteq Q'$ and
minimize the generation of equivalent queries.
In order to do this, we can use three of the four inverse operations described for the previous
phase, i.e., the inverse extension, inverse join and inverse selection operations.
We do not need to use the inverse projection operation since
we do not want those queries that are diagonally contained in $Q$,
but only those queries that are regularly contained in $Q$
as defined in Section~2.
Still, several queries will be generated which are equivalent with
previously generated queries, and hence this should again be tested.

\section{Sample run}\label{sample}

Suppose we are given an instance of the beer drinkers database
used throughout this paper, as shown in Figure~\ref{figex}.

\begin{figure}
\footnotesize
$$
\begin{array}[t]{c}
\mathit{Likes} \\
\begin{array}{|cc|}
\hline
\mathit{Drinker} & \mathit{Beer} \\
\hline
\text{Allen} & \text{Duvel} \\
\text{Allen} & \text{Trappist} \\
\text{Carol} & \text{Duvel} \\
\text{Bill} & \text{Duvel} \\
\text{Bill} & \text{Trappist} \\
\text{Bill} & \text{Jupiler} \\
\hline \end{array}
\end{array}
\qquad
\begin{array}[t]{c}
\mathit{Visits} \\
\begin{array}{|cc|}
\hline
\mathit{Drinker} & \mathit{Bar} \\
\hline
\text{Allen} & \text{Cheers} \\
\text{Allen} & \text{California} \\
\text{Carol} & \text{Cheers} \\
\text{Carol} & \text{California} \\
\text{Carol} & \text{Old Dutch} \\
\text{Bill} & \text{Cheers} \\
\hline \end{array}
\end{array}
\qquad
\begin{array}[t]{c}
\mathit{Serves} \\
\begin{array}{|cc|}
\hline
\mathit{Bar} & \mathit{Beer} \\
\hline
\text{Cheers} & \text{Duvel} \\
\text{Cheers} & \text{Trappist} \\
\text{Cheers} & \text{Jupiler} \\
\text{California} & \text{Duvel} \\
\text{California} & \text{Jupiler} \\
\text{Old Dutch} & \text{Trappist} \\
\hline \end{array}
\end{array}
$$
\caption{Instance of the beer drinkers database.}
\label{figex}
\end{figure}

We now show a small part of an example run of the algorithm presented in the previous section.
In the first phase, all frequent conjunctive queries need to be found,
starting from the most general conjunctive queries.
Let the maximum number of atoms in de body of the query be limited to $2$,
and let the minimal support threshold be $2$,i.e., at least $2$ tuples are needed
in the output of a query to be considered frequent.
Then, the initial set of candidate queries $C_1$, consists of the $6$ queries as shown in Figure~\ref{level1}.
\begin{figure}
\footnotesize
$$
\begin{array}{l}
\hline
  Q_1(x_1,x_2,x_3,x_4) \coloneq \mathit{likes}(x_1,x_2), \mathit{likes}(x_3,x_4) \\
  Q_2(x_1,x_2,x_3,x_4) \coloneq \mathit{likes}(x_1,x_2), \mathit{visits}(x_3,x_4)\\
  Q_3(x_1,x_2,x_3,x_4) \coloneq \mathit{likes}(x_1,x_2), \mathit{serves}(x_3,x_4)\\
  Q_4(x_1,x_2,x_3,x_4) \coloneq \mathit{visits}(x_1,x_2), \mathit{visits}(x_3,x_4)\\
  Q_5(x_1,x_2,x_3,x_4) \coloneq \mathit{visits}(x_1,x_2), \mathit{serves}(x_3,x_4)\\
  Q_6(x_1,x_2,x_3,x_4) \coloneq \mathit{serves}(x_1,x_2), \mathit{serves}(x_3,x_4)\\
  \hline
\end{array}
$$
\caption{Level $1$.}
\label{level1}
\end{figure}
Obviously, the support of each of these queries is $36$, and hence, $F_1 = C_1$.
To generate all candidate conjunctive queries for level $2$, we need to apply
the four specialization operations to each of these $6$ queries.
Obviously, the extension operation is not yet allowed, since this would result in a conjunctive
queries with $3$ atoms in their bodies.
We can apply the Join operation on $Q_1$, resulting in queries $Q_7$ and $Q_8$, as shown in Figure~\ref{level2}.
Similarly, the join operation can be applied to $Q_4$ and $Q_6$, resulting in $Q_9,Q_{10}$ and $Q_{11},Q_{12}$ respectively.
However, the Join operation is not allowed on $Q_2,Q_3$ and $Q_5$, since for each of them, there always exists
a query in which it is contained and which is not yet known to be frequent.
For example, if we join $x_1$ and $x_3$ in query $Q_2$,
resulting in  $Q(x_1,x_2,x_3) \coloneq \mathit{likes}(x_1,x_2), \mathit{visits}(x_1,x_3)$,
then this query is contained in $Q'(x_1,x_2,x_4) \coloneq \mathit{likes}(x_1,x_2), \mathit{visits}(x_3,x_4)$,
of which the frequency is not yet known.  Similar situations occur for the other possible joins on $Q_2,Q_3$ and $Q_5$.
The selection operation can also not be applied to any of the queries, since for each variable we would
select, there always exists a more general query in which that variable is projected, but not selected, and hence,
the frequency of such queries is yet unknown.
We can apply the projection operator on any variable of queries $Q_1$ through $Q_6$,
resulting in queries $Q_{13}$ to $Q_{37}$.
\begin{figure}
\footnotesize
$$
\begin{array}{lr}
\hline
  Q_7(x_1,x_2,x_3) \coloneq \mathit{likes}(x_1,x_2), \mathit{likes}(x_1,x_3) \\
  Q_8(x_1,x_2,x_3) \coloneq \mathit{likes}(x_1,x_2), \mathit{likes}(x_2,x_3) \\
  Q_9(x_1,x_2,x_3) \coloneq \mathit{visits}(x_1,x_2), \mathit{visits}(x_1,x_3) \\
  Q_{10}(x_1,x_2,x_3) \coloneq \mathit{visits}(x_1,x_2), \mathit{visits}(x_2,x_3) \\
  Q_{11}(x_1,x_2,x_3) \coloneq \mathit{serves}(x_1,x_2), \mathit{serves}(x_1,x_3) \\
  Q_{12}(x_1,x_2,x_3) \coloneq \mathit{serves}(x_1,x_2), \mathit{serves}(x_2,x_3) \\
  Q_{13}(x_2,x_3,x_4) \coloneq \mathit{likes}(x_1,x_2), \mathit{likes}(x_3,x_4) \\
  \vdots \\
  Q_{37}(x_1,x_2,x_3) \coloneq \mathit{serves}(x_1,x_2), \mathit{serves}(x_3,x_4)\\
  \hline
\end{array}
$$
\caption{Level $2$.}
\label{level2}
\end{figure}
In stead of showing the next levels for all possible queries, we will show only single path,
starting from query $Q_7$.  On this query, we can now also apply the projection operation on $x_3$.
This results in a redundant atom which can be removed, resulting in the query $Q_7'(x_1,x_2) \coloneq \mathit{likes}(x_1,x_2)$.
Again, for the next level, we can use the projection operation on $x_2$, now resulting in $Q_7''(x_1) \coloneq \mathit{likes}(x_1,x_2)$.
Then, for the following level, we can use the selection operation on $x_2$,
resulting in the query $Q_7'''(x_1) \coloneq \mathit{likes}(x_1,\text{`Duvel'})$.
Note that if we had selected $x_2$, using the constant $\text{`Trappist'}$, then the resulting query
would not have been frequent and would have been removed for further consideration.
If we repeatedly apply the four specialization operations until the levelwise algorithm stops, because
no more candidate conjunctive queries could be generated anymore, the second phase can start generating
confident association rules from all generated frequent conjunctive queries.
For example, starting from query $Q_7'''$, we can apply the inverse selection operation, resulting in $Q_7''$.
Since both these queries have support $3$, the rule $Q_7'' \Rightarrow Q_7'''$ holds with $100\%$ confidence,
meaning that every drinker that likes a beer, also likes Duvel, according to the given database.

\section{Conclusions and future research} \label{conclusion}

In the future, we plan to study subclasses of
conjunctive queries for which there exist efficient
candidate generation algorithms up to equivalence.
Possibly interesting classes are conjunctive queries
on relational databases that consist of only binary relations.
Indeed, every relational database can  be decomposed into a database
consisting of only binary relations.
If necessary, this can be further simplified by only considering
those conjunctive queries that can be represented by a tree.
Note that one of the underlying challenges that always reappears
is the equivalence test, which can be computed efficiently on
tree structures.
Other interesting subclasses are the class of acyclic conjunctive queries
and queries with bounded query-width, since also for these structures,
equivalence testing can be done efficiently~\cite{acyclic}.

However, by limiting the search space to one of these subclasses,
Proposition~\ref{whycontain} is no longer valid, since the intersection
of two queries within such a subclass does not necesserally result in a
conjunctive query which is also in that subclass.

Another important topic is the improvement of performance issues
for evaluating all candidate queries.
Also the problem of allowing flexible constraints
to efficiently limit the search space to an interesting subset of all
conjunctive queries, is an important research topic.


\end{document}